\documentclass[useAMS,usenatbib]{mn2e}
\usepackage{graphicx}

\def\ucr{${\rm cts\,s^{-1}}$ }
\def\ucre{${\rm cts\,s^{-1}}$}
\def\ergs{${\rm erg\,s^{-1}\,cm^{-2}}$ }
\def\ergse{${\rm erg\,s^{-1}\,cm^{-2}}$}
\def\ulum{${\rm erg\,s^{-1}}$ }
\def\ulume{${\rm erg\,s^{-1}}$}
\def\nh{$N_{\rm H}$ }
\def\nhe{$N_{\rm H}$}
\def\nhi{$N_{\rm H~I}$ }

\def\nhex{$N_{\rm H}^{\rm exc}$ }

\def\unh{${\rm cm^{-2}}$ }
\def\unhe{${\rm cm^{-2}}$}

\def\chisq{$\chi^2$ }

\def\gnh{$N_{\rm H}^{\rm Gal}$ }
\def\gnhe{$N_{\rm H}^{\rm Gal}$}
\def\rxj{RX\,J1028.6-0844 }
\def\rxje{RX\,J1028.6-0844}
\def\gb{GB\,1428+4217 }
\def\gbe{GB\,1428+4217}
\def\pmn{PMN\,J0525-3343 }
\def\pmne{PMN\,J0525-3343}

\def\rosat{{\it ROSAT} }
\def\rosate{{\it ROSAT}}
\def\asca{{\it ASCA }}
\def\ascae{{\it ASCA}}
\def\sax{{\it BeppoSAX }}
\def\saxe{{\it BeppoSAX}}
\def\xmm{{\it XMM-Newton }}
\def\xmme{{\it XMM-Newton}}
\def\chandra{{\it Chandra }}
\def\chandrae{{\it Chandra}}
\def\mnras{\rm MNRAS}
\def\apj{\rm ApJ}
\def\apjl{\rm ApJL}
\def\apjs{\rm ApJS}
\def\aj{\rm AJ}
\def\aap{\rm A\&A}

\title[XMM observations of the high-redshift quasar RX\,J1028.6-0844]
{XMM observations of the high-redshift quasar RX\,J1028.6-0844 at $z$=4.276:
soft X-ray spectral flattening}
\author[W. Yuan, et al.]{W. Yuan$^{1}$\thanks{Present address:
Yunnan Astronomical Observatory,
Chinese Academy of Sciences,
Phoenix Hill,  PO Box 110,
Kunming, Yunnan, 650011    China; E-mail: wmy@ynao.ac.cn
},
A.C. Fabian$^{1}$,
A. Celotti$^{2}$,
R.G. McMahon$^{1}$,
and
M. Matsuoka$^{3}$\\
$^{1}$University of Cambridge, Institute of Astronomy, 
Madingley Road, Cambridge, CB3 0HA; E-mail: wmy@ast.cam.ac.uk (wy)\\
$^{2}$SISSA, via Beirut 2-4, 34014 Trieste, Italy\\
$^{3}$Japan Aerospace Exploration Agency (JAXA), 
Tsukuba Space Center,  Tsukuba, Ibaraki 305-8505, Japan
}

\begin{document}

\date{Accepted for publication}

\pagerange{\pageref{firstpage}--\pageref{lastpage}} \pubyear{2003}

\maketitle

\label{firstpage}

\begin{abstract}
We present results from a new XMM-Newton observation of the high-redshift
quasar RX\,J1028.6-0844 at a redshift of 4.276.
The soft X-ray spectral flattening, 
as  reported  by a study with ASCA previously 
(Yuan et al.\ 2000, ApJ 545, 625), is confirmed to be present, 
however, with reduced column density when modelled by absorption.
The inferred column density for absorption intrinsic to the quasar
is 2.1$(^{+0.4}_{-0.3})\times\,10^{22}$\,\unh for cold matter, and
higher for ionised gas.
The spectral flattening shows  remarkable similarity
with those of two similar objects,
GB\,1428+4217 (Worsley et al.\ 2004, MNRAS 350, L67) 
and PMN\,J0525-3343 (Worsley et al.\ 2004, MNRAS 350, 207).
The results improve upon 
those obtained from a previous short-exposure observation for 
RX\,J1028.6-0844 with XMM-Newton (Grupe et al.\ 2004, AJ 127, 1).
A comparative study of the two XMM-Newton observations 
reveals a change in the power-law photon index from 
$\Gamma\simeq$1.3 to 1.5 on timescales of about one year.
A tentative excess emission feature in the rest-frame  5--10\,keV
band is suggested, which is similar to that
marginally suggested for GB\,1428+4217.
\end{abstract}


\section{Introduction}
High-redshift quasars are powerful cosmological probes 
to study the evolution of massive black holes and 
quasar environments in the early universe.
Previous X-ray observations suggested the presence of
soft X-ray spectral flattening\footnote{
This effect is commonly referred to as excess absorption
in the literature. Here we use a generalised
term in consideration of possible alternative explanations.
} 
in some radio-loud quasars at
redshifts $z$=2--3 
(Wilkes et al.\ 1992, Elvis et al.\ 1994, Cappi et al.\ 1997,
Fiore et al.\ 1998, Yuan \& Brinkmann 1998, Reeves \& Turner 2000).
This result is strengthened and extended to higher redshifts
by its detection in a few extremely X-ray/radio-loud quasars at $z>$4,
namely \rxj (Yuan et al.\ 2000), 
\gb (Boller et al.\ 2000, Fabian et al.\ 2001b), 
and PMN\,J0525-3343 (Fabian et al.\ 2001a), 
with \rosate, \ascae, and \saxe.
These objects seem to have characteristics typical of blazars
(Fabian et al.\ 1997, 1998; Zickgraf et al.\ 1997,
Moran \& Helfand 1997, Hook \& McMahon 1998).
The most plausible explanation for this effect is
photoelectric absorption of soft X-rays by associated medium
with column densities of $10^{22-23}$\,\unhe, 
although intrinsic spectral flattening cannot be excluded.
The physical implication of this effect has been discussed 
extensively in the literature in terms of excess absorption (Elvis et al.\ 1998,
Yuan et al.\ 2000, Fabian et al.\ 2001a,b) and 
intrinsic  breaks in the X-ray spectra of blazars 
(Fabian et al.\ 2001a,b).

The contemporary X-ray observatories  
\xmm and \chandra should be able to test these previous findings.
Indeed,  soft X-ray spectral flattening
has been confirmed to be present in \gb and \pmn 
(\xmme, Worsley et al.\ 2004a,b),
and in  some other objects at lower redshifts, 
e.g.\ PKS\,2126-0158 at $z$=3.27 
(\xmme, Ferrero \& Brinkmann 2003;
\saxe, Fiore et al.\ 2003). 
Tentative evidence was also found in the combined spectra of
several $z>4$, moderately radio-loud quasars 
(\chandra, Bassett et al.\ 2004).

\rxj was first detected as an X-ray source in the 
\rosat All-sky Survey (RASS) and was identified as a quasar
at $z$=4.276 (Zickgraf et al.\ 1997).
It is also a radio source (PKS\,B1026-084) with
a flux density of 220\,mJy at 5\,GHz (Otrupcek \& Wright 1991)
and a flat radio spectrum.
Its X-ray colors in the \rosat energy band imply
a hard spectrum (Zickgraf et al.\ 1997).
Its first X-ray spectrum, as obtained with a long \asca observation
made in 1999, flattens substantially towards soft X-ray energies
(Yuan et al.\ 2000);
an excess (cold) absorption model required a column density
of $\sim 2 \times 10^{22}$\,\unh for local absorber
or $\sim 2 \times 10^{23}$\,\unh for absorber intrinsic to the quasar.
A later short-exposure \xmm observation found, however, 
only marginal evidence for excess absorption (Grupe et al.\ 2004).
In this paper we report on a new \xmm observation of \rxj
with an exposure much longer  than the previous observation.
The measured X-ray spectrum---with 
substantially improved photon statistics---confirms 
the presence of the soft X-ray spectral flattening as detected
by \ascae.
The observations and data reduction are described in Sect.\,\ref{sect:obs}.
We present the spectral analysis in Sect.\,\ref{sect:xspec},
including a re-analysis of the previous \xmm observation.
Discussion of the results is given in Sect.\,\ref{sect:discuss},
including comparisons with previous observations
and with other similar objects.
Conclusions are summarised in Sect.\,\ref{sect:conclu}.
We adopt $H_0$=71\,km\,s$^{-1}$\,Mpc$^{-1}$,
$\Omega_{\Lambda}$=0.73, and $\Omega_{\rm m}$=0.27.
The Galactic column density in the direction of \rxj is
\gnh$=4.59 \times 10^{20}\,\rm cm^{-2}$ (Dickey \& Lockman 1990).
Errors are quoted at the $1\,\sigma$ level for one parameter of interest
unless stated otherwise.

\section{Observation and data reduction}
\label{sect:obs}
The quasar \rxj  was observed with \xmm on June 13th, 2003
during satellite revolution 643 (observation ID 0153290101).
The EPIC (European Photon Imaging Camera) 
 MOS1, MOS2, and PN cameras were operated in the
`primary full window' imaging mode and a thin filter was used
to screen out optical/UV light. 
The observational log is shown in Table\,1.
The \xmm Science Analysis System (SAS, v.6.0) and 
the most up-to-date calibrations (August 2004)
were used for data reduction.
We followed standard data reduction and screening procedures. 
A fraction of the observation period  suffered from  
high flaring background caused by soft protons. 
By inspecting the light curve of energy $E>$10\,keV, single events 
in the whole field of view, these periods were identified as
having count rates higher than 1\,\ucr and 0.35\,\ucr for
PN and MOS detectors, respectively, as recommended 
by the \xmm Science Operation Centre (SOC). 

   \begin{table*}
   \begin{center}
      \caption[]{Summary of the XMM observations and data reduction
      information for each EPIC detector}
         \label{tab:xmmobs}
         \begin{tabular}{lccc}
            \hline 
            \noalign{\smallskip}
         &  PN     &   MOS1   & MOS2 \\
            \noalign{\smallskip}
            \hline
            \noalign{\smallskip}
observation duration (ks) & 41.5 & 43.1 & 43.1 \\
good exposure (ks)  & 15.8 & 21.3 & 22.1  \\
energy band used (keV)   & 0.2--10 & 0.3-10 & 0.3--10 \\
events pattern used & 0--4  & 0--12 & 0--12\\
source extraction radius  & 32\arcsec & 32\arcsec & 32\arcsec\\
source+BGD counts     & 5024  & 2006 & 2080 \\
net source counts      & 4854  & 1920 & 1986 \\
source count rate ($10^{-2}$\ucre) &30.7 $\pm$ 0.5 & 9.0$\pm$0.2 & 9.0$\pm$0.2\\
flux$^a$ ($10^{-12}$\ergs) & 1.16$\pm$0.04 & 1.14$\pm$0.07 &  1.14$\pm$0.06 \\ 
flux$^b$ ($10^{-12}$\ergs) & 1.17$\pm$0.04 & 1.15$\pm$0.07 &  1.15$\pm$0.06 \\ 
            \noalign{\smallskip}
            \hline 
         \end{tabular}
\begin{list}{}{}
\item[$^{\mathrm{a}}$] Flux in the 1--10\,keV band; the model of
a power-law with intrinsic absorption at $z$=4.276 plus \gnh is used.
\item[$^{\mathrm{b}}$] Galactic absorption corrected 
flux in the 1--10\,keV band; the same model as in $a$ is used.
\end{list}
\end{center}
\end{table*}

The quasar was detected at a sky position 
RA=10$^h$\,28$^m$\,38$^s$.84, Dec=$-08^o$\,44\arcmin\,38\arcsec.3
(J2000), 0\arcsec.6 away from the position of its radio counterpart
(Simbad database)
and 2\arcsec.1 of its optical counterpart (Zickgraf et al.\ 1997). 
Source X-ray events  were extracted from a circle of 32\,\arcsec radius,
which corresponds to the $\simeq$\,87 per cent encircled energy radius.
Background events were extracted from
source-free regions using a concentric annulus of 52/128\,\arcsec
radii for the MOS detectors,
and circles of  32\,\arcsec radius at 
the same CCD read-out column as the source position for the PN detector.
X-ray images, light-curves, and spectra were generated from 
the extracted, cleaned events for the source and background.
Parameters for data screening and source extraction
are listed in Table\,1.
No photon `pile-up' problem was found, as expected.
nor effect of low-energy noise above  0.2\,keV.
We used the spectral range of 0.2--10\,keV for PN and 0.3--10\,keV for MOS.
The EPIC response files ({\it rmf} and {\it arf}) were generated
using the source  information on the detectors. 
The EPIC spectra were re-binned to have a 
minimum of 30 counts in each bin. 

The source profile in the 0.3--2\,keV band was compared against 
the point spread function of the detectors
(FWHM of 5\,\arcsec for MOS and 6\,\arcsec for PN)
and was  found to be consistent with a point-like source.
There was no significant variability found during the 40\,ks exposure, 
though a $\simeq$10 per cent drop in count rates (averaged over $\sim$5\,ks) 
was marginally detected,
 from 0.316$\pm$0.09\,\ucr at the beginning  to 0.271$\pm$0.08\,\ucr 
towards the end of the observation.

\section{X-ray spectral analysis}
\label{sect:xspec}
Due to an increase in the surface charge loss properties of the CCDs,
which degrades the energy resolution, there has been a
time-dependent, significant change in the low-energy redistribution 
properties of the MOS cameras. 
This effect has been taken into account in the most up-to-date MOS
calibration files; however, some small systematic uncertainty may
still remain  at below 0.5\,keV 
(Kirsch 2004, Kirsch et al.\ 2004).
To minimise possible biases induced in the results, 
we treat the MOS spectra in two ways and compare the results. 
The first was to simply omit the spectral range below 0.5\,keV;
the second was to use the spectral range down to 0.3\,keV and 
introduce a systematic error of 2 per cent in the 0.3--0.5\,keV band
(as recommended in Kirsch 2004, Kirsch et al.\ 2004).
As seen in Table\,\ref{tab:spec_fits},
these two methods yielded statistically consistent results. 
We thus formally quote the results obtained using the
0.3--10\,keV band.
XSPEC (v.11.3) was used for spectral fitting.

\subsection{Soft X-ray spectral  flattening}
\label{sect:specfltt}
\subsubsection{Power-law model with local cold absorption}
We began by fitting 
the spectra of each detector individually
with a single power-law model
\footnote{
$F_{\rm pho} (E) \propto E^{-\Gamma}$, where $\Gamma$ is the photon index.}
modified by neutral absorption with a column density \nh as 
a free parameter.
This model gave acceptable\footnote{
We regard a model fit as acceptable if the 
null hypothesis probability derived from the fit 
is greater than 10 per cent ($P_{\rm null}>$0.1), 
as is commonly quoted.
} 
fits to the PN and the MOS1 spectra (see Table\,\ref{tab:spec_fits}),
but not to the  MOS2 spectrum 
[reduced $\chi^2$=1.4 for 53 degree of freedom (d.o.f.),
i.e.\ a null hypothesis probability of $P_{\rm null}$=0.02 only]. 
Inspection of the $\chi^2$ residuals of the fit singled out 
one energy bin at 1.82\,keV (a width of 80\,eV), 
which contributed 11 out of the total $\chi^2$ of 76. 
The feature can be fitted with either a narrow notch feature
at $E$=1.83$\pm$0.02\,keV (a width of 49$^{+55}_{-19}$\,eV
and a covering fraction 0.99$^{+0.01}_{-0.53}$)
or a Gaussian absorption line ($E$=1.83\,keV) of infinitely small width.
This energy corresponds to 9.7\,keV in the quasar's rest-frame,
at which no known physical absorption feature is present. 
On the other hand, it is coincident with the 
instrumental absorption feature at 1.84\,keV due to the Silicon edge.
Furthermore, it appears in neither the PN nor the MOS1 spectrum.
Excluding the 1.82\,keV bin reduced  the \chisq by 12, 
and resulted in an acceptable fit ($\chi^2$=1.2 for 52 d.o.f.)
with fitted parameters  in good agreement with those for MOS1
within 1-$\sigma$ errors.
We therefore consider the disagreement between MOS2 and
MOS1/PN to be due to inadequate MOS2 calibration around the Silicon edge,
and ignore this energy bin hereafter.
Parameters (including normalisations) for the two MOS spectra 
were tied together to be the same in joint fitting (MOS1+2).
The fitted \nh of $\sim 1.1 \times\,10^{21}$\,\unhe) 
is significantly higher than \gnh 
(Table\,\ref{tab:spec_fits}). 
The photon index is now $\Gamma\simeq$1.55, typical of blazars.

A power-law with fixed Galactic absorption
(4.59\,$\times\,10^{20}$\,\unhe)  yielded unacceptable fit
and a flat photon index ($\Gamma$=1.3--1.4, Table\,\ref{tab:spec_fits}).
The improvement in \chisq for the fit with freely fitted \nh 
over that with fixed \nhe=\gnh is substantial---$\chi^2$ 
was reduced by  36 for PN and by 31 for a joint MOS1+2 fit 
for one additional free parameter. 
Applying the $F$-test\footnote{\label{fn:ftest} 
Recently, Protassov et al.\ (2002)
have questioned the validity of using the $F$-test to test for
the significance of adding an additional spectral component,
as this involves testing a hypothesis that is on the boundary of
the parameter space. 
We note that the $F$-test we used here is not affected
by the boundary condition problem, 
because we are testing for the significance of the different values in \nh 
between \nhe=\gnh (null hypothesis)
and the free-fitted value,
rather than for the addition of a model component.
}
 (Bevington \& Robinson 1992)
gives a chance probability $\ll$0.001.

\begin{figure}
\includegraphics[angle=-90,width=\hsize]{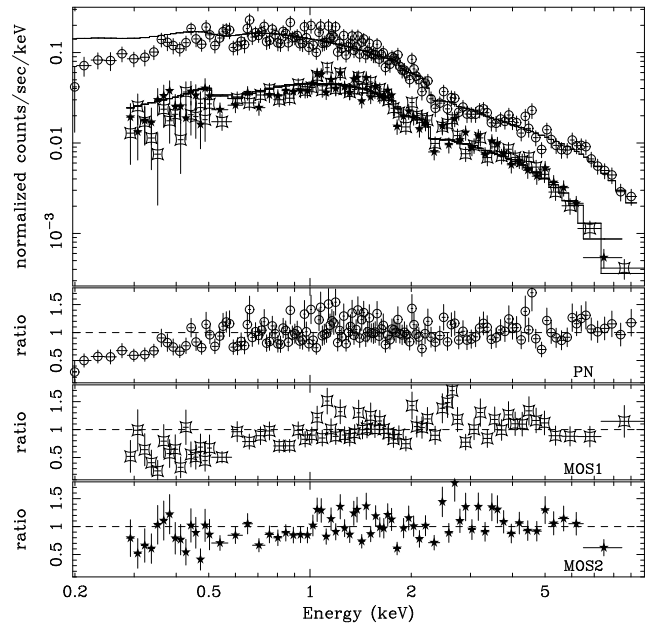}
 \caption{\label{fig:fit_extrp}
The spectra of the PN (circles), MOS1 (squares), and MOS2 (stars)
cameras and the residuals as data-to-model ratio.
The model is the best-fit power-law (with Galactic absorption) 
to the joint PN+MOS1+MOS2 spectra
within the restricted 1--10\,keV energy band and is extrapolated
down to the low energies.
A systematic deviation from a power-law model with Galactic absorption
towards low energies (below 1\,keV) is clearly indicated.
}
\end{figure}

\begin{figure}
\includegraphics[angle=-90,width=\hsize]{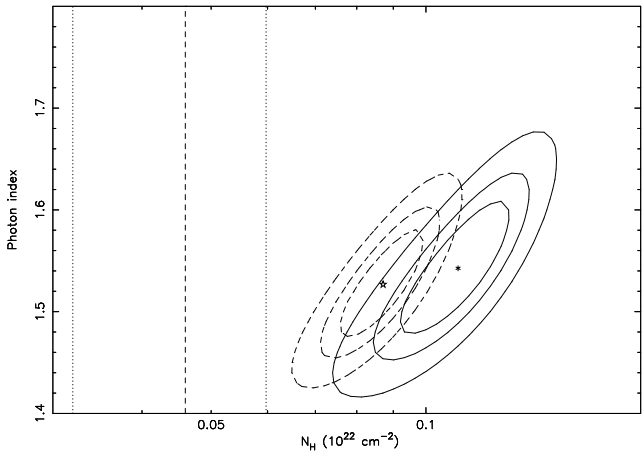}
 \caption{\label{fig:cont_nh_gam}
Confidence contours of fitted  total column density and photon index
for the model of a power-law with local neutral absorption. 
The contours are at the 68, 90, and 99 per cent confidence, respectively, 
for two  interesting parameters.
Solid contours: the joint MOS spectra in 0.3--10\,keV;
dashed contours: the PN spectrum. 
Also indicated by lines are the Galactic column density (dashed) 
and its conservative 30\% uncertainty range (dotted).
Absorption in excess of the Galactic value is evident for such a model.
}
\end{figure}

For fixed  \nhe=\gnhe,
acceptable fit was obtained only within the restricted 1--10\,keV range.
We plotted in Fig.\,\ref{fig:fit_extrp} the data,
the best-fit model (for a joint PN and MOS1+2 spectral fit) 
and its extrapolation down to the low-energy end
of the detectors, and the data-to-model ratio as residuals. 
A systematic deficit of photons below 1\,keV is clearly indicated.
Fig.\,\ref{fig:cont_nh_gam} shows the confidence contours for
the free-fitted total \nh (Galactic plus excess)  and $\Gamma$
for PN and MOS1+2, respectively.
Absorption in excess of \gnh is evident.
It is noted that PN gives systematically lower \nh values than MOS.
We regard the results from MOS to be more reliable than those from PN
in consideration of the EPIC cross calibration uncertainties
as discussed in details in Appendix\,\ref{sect:newarf}.

\begin{table*}
\caption{Results of X-ray spectral fits 
\label{tab:spec_fits}}
\begin{tabular}{lcccl}
 \hline 
            \noalign{\smallskip}
Detector & \nhe$^{a}$ & $\Gamma$ & $\chi^2$/dof & $P_{\rm null}^{b}$\\  \hline
\multicolumn{5}{l}{\it local neutral absorption, free \nh} \\  \noalign{\smallskip}
PN$^c$         & 8.7$\pm$0.7  & 1.53$\pm$0.03 & 137.8/147 & 0.69\\
MOS1 (0.5--10keV)   & 12.8$\pm$2.5 & 1.54$\pm$0.06 & 57.7/52   & 0.27 \\
MOS1 (0.3--10keV)   & 12.7$\pm$2.0 & 1.54$_{-0.03}^{+0.06}$ & 68/67   & 0.44 \\
MOS2 (0.5--10keV)   & 11.3$\pm$2.5 & 1.57$_{-0.03}^{+0.06}$ & 63.7/52 & 0.13\\
MOS2 (0.3--10keV)   & 9.0$\pm$1.6 & 1.53$\pm$0.05 & 76/67 & 0.21\\
MOS1+2 (0.5--10keV) & 12.1$\pm$1.7 & 1.56$_{-0.02}^{+0.05}$ &122.8/107 & 0.14 \\
MOS1+2 (0.3--10keV) & 11.1$\pm$1.2 & 1.54$\pm$0.04 &  150/137 & 0.21\\ 
\hline
\multicolumn{5}{l}{\it local neutral absorption, fixed \nh} \\ \noalign{\smallskip}
PN         & 4.59 (fix) & 1.37$\pm$0.02 & 173/148 & 0.07\\
MOS1 (0.5--10keV)   & 4.59 (fix) & 1.36$\pm$0.04 & 69/53 & 0.07\\
MOS1 (0.3--10keV)   & 4.59 (fix) & 1.30$\pm$0.03 & 91/68 & 0.03\\
MOS2 (0.5--10keV)   & 4.59 (fix) & 1.43$\pm$0.04 & 71/53 & 0.05\\
MOS2 (0.3--10keV)   & 4.59 (fix) & 1.40$\pm$0.03 & 84/68 & 0.09\\
MOS1+2 (0.5--10keV) & 4.59 (fix) & 1.40$\pm$0.03 & 141/108 &0.017\\
MOS1+2 (0.3--10keV) & 4.59 (fix) & 1.35$\pm$0.02 &  181/138 & 0.008 \\
\hline
\multicolumn{5}{l}{\it power-law + \gnh and excess absorption at
            $z=4.276$}\\ \noalign{\smallskip}
Detector  & \nhex         & $\Gamma$  & $\chi^2$/dof &$P_{\rm null}$\\  \noalign{\smallskip}
PN                    & 149$\pm$30 & $1.49\pm0.03$ & 136/147 &\\
MOS1+MOS2 (0.5--10keV)& $393^{+99}_{-70}$ & $1.53^{+0.04}_{-0.02}$ & 122/107 &0.15\\
MOS1+MOS2 (0.3--10keV)& 222$\pm$46 & 1.49$\pm$0.03 &153/137 & 0.17\\
 \hline
\multicolumn{5}{l}{\it broken power-law + Galactic absorption
            (PN+MOS1+2) }\\ \noalign{\smallskip}
    & $E_{\rm break}^{d}$ &  $\Gamma_{\rm low-E}$ & $\chi^2$/dof  &$P_{\rm null}$\\ \noalign{\smallskip}
PN  &$0.50^{+0.04}_{-0.03}$ & $-1.55^{+0.70}_{-}$ & 133/146 & 0.77 \\
MOS1+2 (0.5--10keV) &$1.1^{+0.1}_{-0.1}$  &$0.86^{+0.17}_{-0.19}$ & 122/105& 0.12\\
MOS1+2 (0.3--10keV) &$1.1^{+0.2}_{-0.1}$ &$0.92^{+0.10}_{-0.23}$ & 148/135& 0.21\\
 \noalign{\smallskip} \hline 
\end{tabular}
\begin{list}{}{}
\item[a]{Column density of hydrogen in units of $10^{20}$\,\unh}
\item[b]{Null hypothesis probability of the \chisq test for the fit}
\item[c]{The \nh values from PN are possibly under-estimated;
see Sect.\,\ref{sect:disc_xmm} 
and Appendix\,\ref{sect:newarf} for discussion.}
\item[d]{The break energy of broken power-law in keV. 
When no errors are given, the value is unconstrained within a physically
meaningful range.}
\end{list}
\end{table*}

\subsubsection{Absorption intrinsic to the quasar} 
\begin{figure}
\includegraphics[angle=-90,width=\hsize]{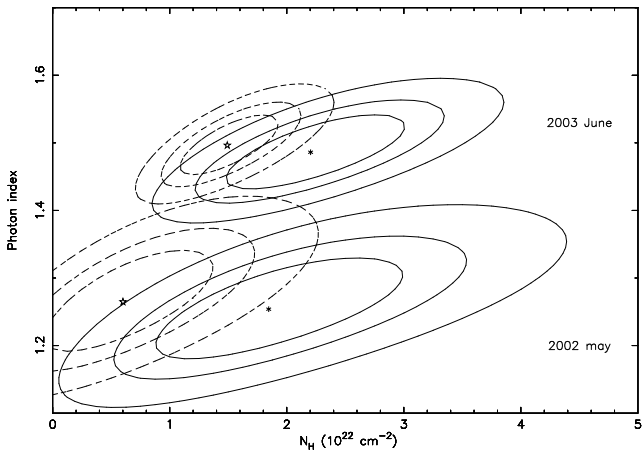}
 \caption{\label{fig:cont_gamnh_2obs}
Column density of intrinsic absorber versus power-law photon index 
as measured with MOS (solid) and PN (dashed) detectors.
The absorber is assumed to be at  the quasar redshift 4.276
and to have cosmic abundances.
The contours correspond to confidence levels of 68, 90, and 99 per cent.
The results from the current (2003) and 
the previous XMM observations (2002) are plotted.
The variation in the spectral slope is significant,
while there is no change in the absorption column density.
      }
\end{figure}

This is the most plausible postulate 
in consideration of the statistical argument presented in
previous work (see Introduction for references).
For neutral absorber, \nhe$\simeq 2\times 10^{22}$\,\unh was found
 assuming cosmic abundances. 
The confidence contours for the excess \nh versus $\Gamma$
are shown in Fig.\,\ref{fig:cont_gamnh_2obs} for  PN (dashed)
and MOS1+2, respectively.
If the metallicity of the absorber is lower than the cosmic value,
which is not unexpected at such a high redshift---could 
be about 10 per cent or less
(e.g.\ Lu et al.\ 1996, Pettini et al.\ 1997, Prochaska \& Wolfe 2000),
the \nh would be correspondingly higher than the value give here.
The absorption-corrected luminosity in the quasar rest-frame
is 9.2\,$\times 10^{46}$\,\ulum  in 1--10\,keV 
and 2.64\,$\times 10^{47}$\,\ulum  in 1--50\,keV, respectively.

If the absorber is close enough to the central source,
the gas is likely to be ionised. 
Indeed, the optical spectra of \rxj taken by Peroux et al.\ (2001) and
Zickgraf et al.\ (1997) show no significant Lyman limit absorption
at 912\,\AA\,.
An estimate of the optical depth of  the  Lyman limit absorption 
places an upper limit on the neutral hydrogen column density 
to be \nhi$\la10^{-17}$\,\unh along the line-of-sight to the optical--UV 
emission region (Yuan et al.\ 2000). 
If the X-ray absorber also intercepts the optical--UV light,
at least moderate ionisation of the gas is required.
The lack of strong optical--UV extinction can be explained
by ionised, dust-free absorber.
It is interesting to note that such an optical--UV property seems 
to be common among this type of objects 
(Yuan et al.\ 2000, Fabian et al.\ 2001a,b).
We tried to model the excess absorption with ionised absorption
({\it absori} in {\it XSPEC}).
The ionisation parameter (as defined in Done et al.\ 1992), however,
could not be constrained with the current data
ranging from an almost neutral to highly ionised absorber. 
The results of such an analysis are given in Fig.\,\ref{fig:cont_xi_nh}, 
where the data from the two \xmm observations are
combined in order to achieve a better constraint on the parameters
(see Sect.\,\ref{setc:fit2obs}).
With increasing ionisation state, the  required column density 
increases from $\sim 2\times 10^{22}$ to $\sim 1\times 10^{23}$\,\unhe.

No redshifted K-shell absorption edges from iron ions 
($E_{\rm edge}$=7.1--9.3\,keV in the rest-frame from 
Fe$_{\rm I}$ to Fe$_{\rm XXVI}$)  are detected.
Assuming cosmic abundance of iron ($3.4\times 10^{-5}$), 
a K-shell edge with optical depth $\tau \sim$ 0.02--0.1
is predicted from the above \nh range for 
neutral--highly ionised iron.
This is consistent with the upper limit on $\tau$ estimated
from the joint PN and MOS1+2 spectra, which ranges from 0.1 to 0.2 
(90 per cent level)
for ions from Fe$_{\rm I}$ to Fe$_{\rm XXVI}$ 
(absorption cross section from
$3.8\times 10^{-20}$ 
to $3.3\times 10^{-20}$\,cm$^2$\,atom$^{-1}$).

\subsubsection{Intrinsic spectral break}
We also considered the possibility that the
soft X-ray spectral flattening is an intrinsic  feature. 
The physical implication of a break in the intrinsic X-ray spectra
of blazars has been discussed in Fabian et al.\ (2001a,b)
in the context of a cut-off in the  energy distribution of 
electrons or  of soft seed photons for Compton scattering.
We modelled the spectral flattening with, as an approximation,
 a broken power-law  modified by local absorption. 
Acceptable fits could be obtained for both PN and MOS1+2.
However, \nh and the low energy photon index 
$\Gamma_{\rm low-E}$ could not be constrained due to strong coupling.
We thus fixed \nhe=\gnhe.
The fitted high-energy photon indices $\Gamma_{\rm high-E}$ are
1.44$\pm$0.04 (PN) and 1.49$\pm$0.06 (MOS1+2), and
the break energy $E_{\rm break}$ and low-energy index $\Gamma_{\rm low-E}$
are listed in Table\,\ref{tab:spec_fits}.
This model gave acceptable fits which are statistically indistinguishable
 from the models of power-law with either local or 
intrinsic absorption.

\subsection{Comparison with a previous observation and spectral variability}
\label{sect:specvary}
\subsubsection{A previous \xmm observation}
\rxj was previously observed with \xmm in revolution 445
with a short exposure of 7\,ks in May, 2002
(PI.\ S.\ Mathur, observation ID: 0093160701).
The results, as published in Grupe et al.\ (2004), 
gave absorption \nh values similar to what we find here,
but, a flatter spectral slope of $\Gamma\simeq 1.3$.
In order to achieve a self-consistent comparison of the two
observations---free from the effects introduced by 
different versions of the evolving calibration and 
data processing software, 
we also analysed the data from that observation.
The data were taken from the \xmm science archive.
The observation was described in Grupe et al.\ (2004).
We used exactly the same data screening criteria and 
source/background extraction regions as used for the current
observation (see Sect.\,\ref{sect:obs} and Table\,\ref{tab:xmmobs}).
The good exposure and source count rate
are 3.7\,ks and 0.39\,\ucr for PN (0.2--10\,keV), 
and  6.9/7.0\,ks and 0.12/0.13\,\ucr for MOS1/2 (0.3--10\,keV).
A comparison with
Table\,\ref{tab:xmmobs} reveals that the broad-band count rates
were higher in the 2002 observation than in 2003 by about 30 per cent.

The PN and MOS spectra were binned to have a minimum of
25 and 20 counts in each bin, respectively.
The results of the spectral fits are in good agreement with 
those obtained by Grupe et al.\ (2004).
For an absorbed power-law model,  
the total local absorption \nh is 
$10.1(^{+2.0}_{-1.2})$\,$\times 10^{20}$\,\unh 
for joint MOS1 and MOS2 (MOS1+2) 
spectra and $6.0(^{+1.3}_{-1.0})$\,$\times 10^{20}$\,\unh for PN.
Again, the fitted \nh is systematically lower for PN than for MOS,
as found in the 2003 observation.
While the absorption \nh is in good agreement
between the two observations, 
the  photon indices are not.
The  photon indices obtained for the 2002 observation
are $\Gamma_{\rm 2002}$=$1.30\pm0.06$ and  $1.27\pm0.05$ 
for MOS1+2 and PN, respectively, i.e.
the spectrum was steeper during the observation of 2003
($\Gamma_{\rm 2003}=1.53\pm0.03$).
This result can be seen clearly in Fig.\,\ref{fig:cont_gamnh_2obs}, 
where excess (neutral) absorption \nh is plotted 
versus  $\Gamma$ for the 2002 observation, 
assuming absorption is intrinsic to the quasar at $z$=4.276.
The spectral steepening remains even when 
only the hard band 2--10\,keV spectra 
were considered ($\Gamma_{\rm 2002}$=$1.33^{+0.08}_{-0.14}$  and 
$1.23^{+0.08}_{-0.13}$ for MOS1+2 and PN, respectively).
The Galactic absorption corrected
flux in the 1--10\,keV band is
1.9\,$\times 10^{-12}$\,\ergs (averaged MOS1 and MOS2 value).

\subsubsection{Joint spectral fit of the two observations}
\label{setc:fit2obs}
We quantified the spectral variability by fitting jointly
the spectra of the two observations.
We used the MOS spectra only,
in consideration of possible PN calibration uncertainties
(see Appendix\,\ref{sect:newarf}).
For MOS1 and MOS2 spectra from the {\em same} observation,
all parameters (including normalisation) were tied together. 
The results are summarised in Table\,\ref{tab:fit_2obs}. 

Firstly, since the fitted \nh of the two observations are in good agreement,
we assumed that there was no variability in the absorption.
Thus the \nh values for the two {\em observations}
were tied together in the fitting.
Absorbed power-law models were used. 
As a test, we tied $\Gamma_{\rm 2002}$=$\Gamma_{\rm 2003}$ together
in the fitting, which resulted in a fit only marginally acceptable
($\chi^2$=242 for 219 d.o.f.).
Setting the two indices as independent parameters improved 
the fit significantly, reducing
$\chi^2$ by 24 for one additional free parameter;
the $F$-test [Bevington \& Robinson 1992;
see Footnote\,\ref{fn:ftest} for
the argument for the validity of the  $F$-test used here
regarding the boundary condition warning discussed by Protassov et al.\ (2002).]
gives a chance probability $\ll 0.001$ for 
$\Gamma_{\rm 2002}$ and $\Gamma_{\rm 2003}$ being the same.
The fit is good ($\chi^2$=217 for 218 d.o.f.),
indicating that the spectra of the two observations can be fitted 
well with two different continua (in slope and normalisation)
attenuated by the same amount of absorption.
The fluxes in  the 0.2--1\,keV band were comparable in the two
observations, while in the 1--10\,keV band 
it was higher by a factor of two in May 2002 compared to June 2003
(Table\,\ref{tab:fit_2obs}). 
The two power-law continua, before attenuation by any absorption,
cross over at $\simeq0.4$\,keV in the observer's frame,
i.e.\ $\simeq2$\,keV in the quasar rest-frame.

\begin{figure}
\includegraphics[angle=-90,width=\hsize]{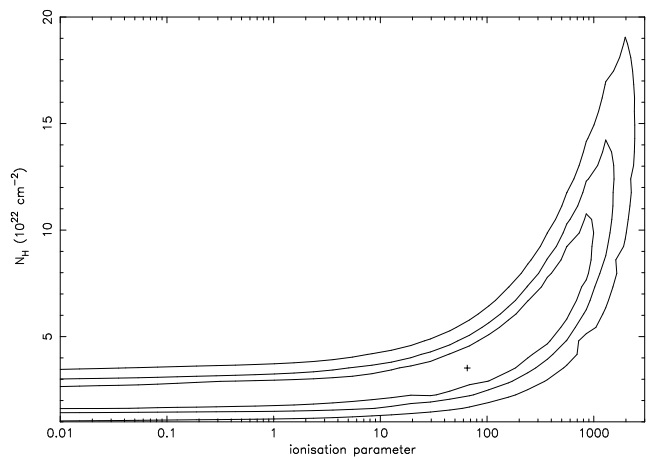}
 \caption{\label{fig:cont_xi_nh}
Confidence contours (at the 68, 90, and 99 per cent level)
for the ionisation parameter and column density
of the ionised absorber model.
The result is obtained from a joint spectral fit to the
MOS spectra from the two \xmm observations in 2003 and 2002
(see Sect.\,\ref{setc:fit2obs}).}
\end{figure}

The combined data set improves the spectral photon statistics
and gives a better constraint on the excess absorption.
For neutral absorber intrinsic to the quasar,
\nh=2.1($^{+0.4}_{-0.3}$)\,$\times 10^{22}$\,\unhe, i.e.\
excess absorption is evident.
An ionised absorber model also gives a good fit,
suggesting that the ionisation status is unconstrained.
The confidence contours for the ionisation parameter and
\nh are shown in Fig.\,\ref{fig:cont_xi_nh}.

Secondly, we tested whether the variation in  spectral slope
could result from variability in the absorption,
such as from the ionisation parameter.
We fitted the ionised absorber model with independent ionisation
parameters and \nhe, but tied the indices together 
($\Gamma_{\rm 2003}$=$\Gamma_{\rm 2002}$).
No satisfactory result could be obtained for this model;
the best fit was significantly worse than that with
the above variable-$\Gamma$ model ($\Delta\chi^2=14$).
This  is not surprising, as the difference in $\Gamma$
arises primarily  in the hard 1--10\,keV band, 
which corresponds to 5.3--53\,keV in the quasar rest-frame.
At such high energies the amount of absorption of X-ray photons is 
decreasing dramatically.

We also fitted a broken power-law to the data
with absorption fixed at \gnhe.
There is a marginal indication of a 
higher break energy 1.6$\pm$0.2\,keV 
for the flatter spectrum in May 2002 than 
for the steeper spectrum in June 2003; 
however, the significance is low.
The low-energy photon indices and the normalisations at 1\,keV
(below or close to the break energies) are comparable 
in the 2002 and 2003 observations.

No spectral variability is detected within a 40\,ks 
duration in the June 2003 observation with \xmme. 

\begin{table}
   \center
      \caption[]{Joint fits to the current (2003) and 
a previous (2002) \xmm observations.
Only the MOS spectra were used.
The absorption \nh is assumed to be the same in the two observations  
and the photon index is freely fitted.
}
         \label{tab:fit_2obs}
         \begin{tabular}{lcc}
            \hline 
            \noalign{\smallskip}
parameter   & June 2003 & May 2002 \\
            \noalign{\smallskip}
            \hline
            \noalign{\smallskip}
\multicolumn{3}{l}{{\it power-law with neutral absorption} ~~~~$\chi^2$/d.o.f.=217/218}\\
photon index & 1.53$\pm$0.04 & 1.31$\pm$0.04 \\
normalisation @1keV ($10^{-4}$) & 1.77$\pm$0.07 & 2.13$\pm$0.09 \\
total absorption \nh ($10^{22}$) & 
\multicolumn{2}{c}{0.11$\pm$0.01 (tied)}  \\
flux$^{a}$ 0.2--1/1--10\,keV ($10^{-13}$) & 1.2/11.5 & 1.4/19.1  \\ 
\noalign{\smallskip}

\multicolumn{3}{l}{{\it neutral absorption at z=4.276, fixed \gnh} $\chi^2$/d.o.f.=221/218}\\
photon index & 1.48$\pm$0.03 & 1.26$\pm$0.04 \\
normalisation @1keV ($10^{-4}$) & 1.65$\pm$0.05 & 1.99$\pm$0.07 \\
excess absorption \nh ($10^{22}$)& \multicolumn{2}{c}{2.1$^{+0.4}_{-0.3}$ (tied)} \\
flux$^{b}$ 0.2--1/1--10\,keV ($10^{-13}$) & 1.7/11.6  & 1.8/19.2  \\ 
luminosity$^{c}$ 1--10/1--50\,keV ($10^{47}$) & 0.92/2.58 & 1.06/3.88  \\ \noalign{\smallskip}

\multicolumn{3}{l}{{\it broken power-law, fixed \gnh} ~~~~$\chi^2$/d.o.f.=214/215}\\
break energy (keV) & 1.1$\pm$0.1    & 1.6$\pm$0.2 \\
low-energy index   & 0.92$\pm$0.11  & 0.87$\pm$0.09 \\
high-energy index  & 1.50$\pm$0.04  & 1.33$\pm$0.07 \\
normalisation @1keV ($10^{-4}$) & 1.54$\pm$0.06 & 1.75$\pm$0.06 \\
\hline
\end{tabular}
\begin{list}{}{}
\item[$^{\mathrm{a}}$] Galactic absorption uncorrected flux in units
of \ergse.
\item[$^{\mathrm{b}}$] Galactic absorption corrected flux in units of \ergse.
\item[$^{\mathrm{c}}$] Absorption corrected  luminosity in the
quasar rest-frame in units of \ulume.
\end{list}
\end{table}

\section{Discussion}
\label{sect:discuss}
\subsection{The presence of soft X-ray spectral flattening}
\subsubsection{\xmm results}
\label{sect:disc_xmm}
We have shown the presence of soft X-ray spectral flattening
in the z=4.276 quasar \rxj using an observation 
made with \xmme.
The result confirms the previous report based on ASCA data 
(Yuan et al.\ 2000).
In the excess absorption scenario,
the derived absorption \nh 
from our analysis are consistent with those
obtained by  Grupe et al.\ (2004) from a previous 
short \xmm observation.
In that study the authors suggested 
that strong excess absorption was marginal. 
This is not surprising given the lower signal-to-noise of their data
and consequently the weaker constraints on \nh compared to 
this work, which benefits from a much longer exposure.

We note that PN tends to give systematically lower \nh values than
MOS. This is most likely due to discrepancies in the 
calibration below 1\,keV between  PN and MOS, 
as reported in the most recent  XMM calibration status 
(Kirsh et al.\ 2004, see also {\it XMM-Newton SOC XMM-SOC-CAL-TN-0018}
\footnote{http://xmm.vilspa.esa.es/es/external/xmm\_sw\_cal/calib/index.shtml}).
Although a complete solution is yet to be reached,
preliminary indications suggest that PN is most likely the cause of
the problem.
A detailed discussion on this issue and a 
quantitative PN--MOS comparison  
taking the most up-to-date calibration into account 
is given in Appendix\,\ref{sect:newarf}.
In summary, we consider the \nh values derived from MOS spectra
to be more reliable.
Furthermore, it may also be the case that 
both PN and MOS yield systematically higher fluxes
at low energies compared to \xmm RGS and \chandrae.
If this turns out to be the case, the true column in the 
absorption scenario could be even higher than what is reported here.

\subsubsection{Comparison with previous results}
The total  \nh assuming local absorption 
is $\simeq (0.11\pm0.01$)\,$\times10^{22}$\,\unh
 measured from the \xmm observations 
(joint MOS from the two observations).
This value is a factor of 2--3 times smaller than the value
measured from ASCA (Yuan et al.\ 2000).
For intrinsic absorption at $z=4.286$, 
the \nh inferred by \xmm (a few times $10^{22}$\,\unh) 
becomes about 10 times smaller than that obtained by \ascae.
Intrinsic variability in the absorption
cannot be ruled out.
However, we speculate that a systematic difference 
in the instrumental calibration of the two missions
might play at least a partial  role.
This is because a  similar trend was also found for \gb and \pmn 
(Worsley et al.\ 2004a,b).
It is not clear which instrument causes the difference.
In the cases of \gb and \pmne, the previous results before \xmm 
were obtained
by a joint fit of {\em both} the \asca and \sax spectra (Fabian 2001a,b).
This fact, together with the aforementioned XMM EPIC calibration
issue, suggests that perhaps both missions, 
rather than merely \xmme, may be the cause.
Improved XMM EPIC calibration and independent investigations by other 
instruments (\chandra or \xmm RGS) are needed to resolve this problem.

Comparing the observations of \xmm in 2003 and of \asca in 1999,
the spectral photon indices are consistent within their mutual
1-$\sigma$ errors; no significant  flux variability ($>$10 per cent)
is detected in the 1--10\,keV band.

\subsubsection{Comparison with other objects}
\begin{figure}
\includegraphics[angle=-90,width=\hsize]{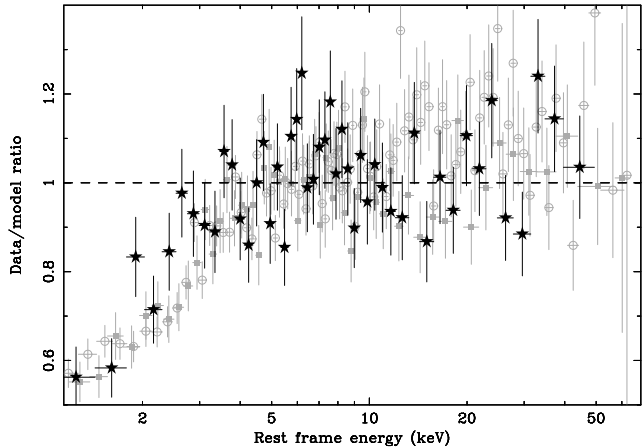}
 \caption{\label{fig:ratio_3obj}
Data-to-model ratio in the rest-frame for \rxj (stars), 
where the model is the best-fit 
power-law with Galactic absorption in the restricted 1--10\,keV
spectral range. 
Only the PN spectrum (May 2003 observation) is plotted, 
which is rebinned for demonstration.
Also plotted are those for  \gb (filled squares) 
and \pmn (open circles) for comparison, 
which were taken from Fig.\,5 in Worsley et al.\ (2004b).
The spectra are plotted in their respective rest-frame energies.
}
\end{figure}

We compare the spectral shape of \rxj with those of \gb and \pmn
in the quasar's rest-frame.
Following Worsley et al.\ (2004b), 
we produced the data-to-model ratio for \rxj 
where the model is the best-fit 
power-law with Galactic absorption in the restricted 1--10\,keV band.
The result is plotted in Fig.\,\ref{fig:ratio_3obj},
together with those for \gb and \pmn from 
Worsley et al.\ (2004b, their Figs.\,2 and 5).
It should be noted that the data-to-model ratio is free from
the effects of instrument response, Galactic absorption, and redshift.
It can be seen that the three spectra agree remarkably well in terms of
the break energy and the shape of the spectral cutoff.
The \nh values of intrinsic (cold) absorbers
of a few times $10^{22}$\,\unhe as measured by \xmm
in these three objects are in good agreement.
The striking spectral similarity shared by these objects 
at different redshifts
argues for a real soft X-ray flattening against instrumental effects,
and suggests a common nature  to this phenomenon.

Another similarity lies in their  optical--UV properties,
which argue for a highly ionised, dust-free absorber model
(Yuan et al.\ 2000; Fabian et al.\ 2001a,b; Worsley et al.\ 2004a,b).
It is interesting to note that excess absorption
in several $z>4$, moderately radio-loud quasars,
as tentatively suggested by their combined \chandra spectra,
has also similar \nh values of a few times  $10^{22}$\,\unh
(Bassett et al.\ 2004). 

\subsection{X-ray spectral variability}

It is worth stressing that while spectral variability
is indeed typical of flat-spectrum quasars,
the extremely flat spectrum in the 2--10\,keV band
during the 2002 observation is consistent only
within the 2\,$\sigma$ uncertainty range
with the limiting value of $\Gamma\simeq 1.5$ for a
relativistic distribution of particles emitting via synchrotron and
inverse Compton (in the simplest hypothesis). 
If the production of such flat high-energy spectra are confirmed, 
revision of the widely accepted emission scenarios would be required.

Although flux variability over short timescales is a distinctive
characteristic of blazar emission, no significant variations 
have been detected within a single observation. 
It should be noted, however, that due to
the high redshift of the source, the intrinsic timescale sampled by
the observation ($\sim$ 2 hr) might be  too short 
to detect  significant variations 
(for powerful quasars doubling timescales of
the order $\sim$several hour to a day might be more typical, 
e.g.\ 3C279, Wherle et al. 1998).

\subsection{Excess emission  around 5--10\,keV?}
Worsley et al.\ (2004b) pointed out possible excess emission
at energies around 5--10\,keV in the quasar rest-frame spectrum
for \gbe. The evidence is only marginal.
Interestingly, the same spectral structure also appears 
in \rxje, as can be seen in Fig.\,\ref{fig:ratio_3obj}.
The similarity of the energy position of this feature in two objects
at different redshifts is remarkable.
If this feature is real, it may come from an additional spectral
component which is peaked around 5--10\,keV in the rest-frame. 
We modelled the \xmm spectra taken in May 2003 by adding a steep
power-law component in the above spectral models, following
Worsley et al.\ (2004b).
Both the PN and MOS1+2 spectra were fitted jointly to improve
the statistics.
A steep  photon index is yielded for the second power-law,
as $\Gamma$=2.3--4.4 for a model with local absorption
and $\Gamma$=1.9--7.2 for  intrinsic absorption at $z$=4.276
(90 per cent confidence range for 1 interesting parameter).
These values are consistent with that obtained for \gbe,
$\Gamma\sim$1.8--2.6 (Worsley et al.\ 2004b).
The model does improve the fit in the 5--10\,keV (rest-frame) band,
though the statistical significance is not high
($\Delta \chi^2 \simeq -7$ for 3 additional free parameters).

If this excess emission feature proves to be real,
it might be the first evidence for the
presence of emission originating from bulk comptonisation on the soft
photon field through which the relativistic jet propagates 
(Begelman \& Sikora 1987). 
The detection of such a feature could carry key clues
to the amount of (cold) leptons flowing in the jet 
(Sikora \& Madejski 2000). 
The lack of its detection in the majority of
objects so far remains a puzzle. 
However, in most cases
the X-ray emission might be dominated by the non-thermal emission from
relativistic particles, and thus the possibility of detecting
such a component could be limited to cases of 
particularly high jet Lorentz factors 
(which would shift its peak up to high energies) 
and/or low-states/steep power-law of the non-thermal 
relativistic component 
(the latter case could of course be tested, in principle).
Observations with even higher signal-to-noise 
than the present ones or stacking spectra from different sources 
might be a way to clarify the issue.

\section{Conclusions}
\label{sect:conclu}
We have presented a new X-ray spectroscopic study of the high redshift
($z$=4.276) quasar \rxj with \xmme.
The high signal-to-noise spectrum confirms the
presence of the soft X-ray spectral flattening, which was
reported previously  with \asca data (Yuan et al.\ 2000).
This spectral feature can be modelled by either excess absorption
of the quasar X-rays or an intrinsic  break at $\sim$1\,keV 
in the X-ray spectra of the source.
In the absorption scenario, the derived  column density for
cold absorber intrinsic to the quasar is 
2.1($^{+0.4}_{-0.3}$)\,$\times 10^{22}$\,\unhe.
This value is comparable to those reported in two similar objects
\gb ($z$=4.72) and \pmn ($z$=4.4) from \xmm observations.
The remarkable similarity in the shape of the spectral cutoff 
among these objects at different redshifts 
argues against instrumental effects as an origin, 
but rather argues for a common nature 
to the soft X-ray flattening in high-$z$ blazars.
In terms of the soft X-ray spectral flattening,
the results are consistent statistically with and
improved upon  those obtained
from a previous short-exposure observation for \rxj
 with \xmm (Grupe et al.\ 2004).
A comparative study of the two \xmm observations 
revealed a spectral steepening from $\Gamma\simeq$1.3 in 2002 to 
$\Gamma\simeq$1.5 in 2003, and a consequent drop in flux
in the hard energy band above $\sim$1\,keV.

The derived columns from \xmm observations, however, are reduced
when compared with the previous \asca results (Yuan et al.\ 2000).
We speculate that this might be due to systematic  instrumental 
effects, probably inherent in both missions.
Future improved \xmm EPIC calibration 
and independent investigations by other 
instruments (such as \chandrae) are needed to resolve this issue.

A tentative excess emission feature in the rest-frame  5--10\,keV
band is suggested, which bares remarkable similarity to that 
marginally imprinted in the X-ray spectrum of
\gb  (Worsley et al.\ 2004b).

\section*{Acknowledgements}
We thank Richard Saxton of the \xmm calibration team
for useful advice on the EPIC calibration issues.
Matt Worsley is thanked for help in
making the plot of Fig.\,\ref{fig:ratio_3obj}.
W.Y. thanks Franz Bauer for comments and 
a careful reading of the manuscript.
ACF thanks the Royal Society for their support.
AC acknowledges the MIUR  and INAF for financial support.
This research has made use of the
NASA/IPAC Extragalactic Database (NED) which is operated by the Jet
Propulsion Laboratory, California Institute of Technology, 
under contract with        
the National Aeronautics and Space Administration.

\appendix

\section{Effect of calibration uncertainty on the results}
\label{sect:newarf}
In the most updated reports on the \xmm calibration status,
Kirsh et al.\ (2004, see also {\it XMM-Newton SOC
XMM-SOC-CAL-TN-0018}
\footnote{http://xmm.vilspa.esa.es/es/external/xmm\_sw\_cal/calib/index.shtml})
demonstrated significant discrepancies at low energies (below $\sim$1keV) 
between the EPIC PN and MOS cameras,
and the EPIC and the RGS\footnote{The Reflection Grating
Spectrometer on-board XMM-Newton.},
as well as overall differences in the \xmm
instruments and those of \chandrae.
While MOS and RGS agree with each other in general,
PN gives higher fluxes below 0.7\,keV
by 10--15 per cent with respect to MOS and 
by 20--30 per cent with respect to RGS (Kirsh et al.\ 2004).
It appears that the problem arises mainly from the
PN calibration at low energies,
although it is not clear whether
the redistribution matrix or the effective area is deficient.
This means that PN spectra tend to give less absorption column
than MOS and RGS,
a trend similar to what we find in this work
(Table\,\ref{tab:spec_fits} and Table\,\ref{tab:fit_2obs}),
as well as those obtained by Grupe et al.\ (2004).
We thus argue that the \nh values derived from MOS spectra
are more reliable.

If the effect is purely due to an effective area problem of PN,
we can test the PN--MOS consistency by 
taking into account the reported difference in the calibration.
To simplify the treatment, we assumed that the PN effective area
below 1\,keV was under-estimated 
by a factor of $f(E)$, which is energy-dependent.
$f(E)$ was estimated from Fig.\,10 in Kirsh et al.\
(2004), in which the MOS/RGS normalisation factors with respect to
PN is plotted versus energies.
We adopted conservative values of  
$f(0.3\,\rm keV)\sim$1.4, $f(0.6\,\rm keV)\sim$1.15, and 
$f(1\,\rm keV)\sim$1.0.
$f(E)$ at any other energies $E$ within 0.3--1.0\,keV
was interpolated using a binomial function.
We corrected the PN effective area (ARF) by multiplying by $f(E)$
at the corresponding energy.
The PN spectral fitting was repeated with is corrected ARF
and the results are listed in Table\,\ref{tab:fit_newarf}.
The fitted total \nh is (11.1$\pm0.8$)\,$10^{20}$\,\unhe,
in excellent agreement with the MOS results,
while $\Gamma$ remains unchanged.
We therefore conclude that the \nh fitted from the PN spectrum is 
likely to be under-estimated. 
More quantitative and reliable estimation of \nh in PN spectra
must await the completion of the PN and MOS 
calibration at the low energies (Kirsh et al.\ 2004).

\begin{table}
\center
\caption{absorbed power-law model fit to the PN spectrum using
corrected effective area ARF calibration}
\label{tab:fit_newarf}
\begin{tabular}{lcccl}
 \hline 
            \noalign{\smallskip}
absorber & \nhe$^{a}$ & $\Gamma$ & $\chi^2$/dof & $P_{\rm null}$\\  \hline
\multicolumn{5}{l}{observation in June 2003} \\ 
local  & 11.1$\pm$0.8  & 1.54$\pm$0.03 & 137/147 & 0.70\\
local  & 4.59 (fix) & 1.29$\pm$0.02 & 218/148 & 1.4$10^{-4}$\\
z=4.276 & 231$\pm$28 & $1.50\pm0.03$ & 134/147 &0.77\\ 
  \hline 
\end{tabular}
\begin{list}{}{}
\item[a]{Column density of hydrogen in units of $10^{20}$\,\unh}
\end{list}
\end{table}


\label{lastpage}

\end{document}